\documentclass[prd,a4paper,10pt,showpacs,preprint,byrevtex,twocolumn]{revtex4} %

\usepackage{amsfonts}
\usepackage{bm,amssymb}
\usepackage{amsmath}

\begin{document}

\title{The few-body problem in terms of correlated gaussians}

\author{Bernard \surname{Silvestre-Brac}}
\email[E-mail: ]{silvestre@lpsc.in2p3.fr} \affiliation{Laboratoire
de Physique Subatomique et de Cosmologie, CNRS/IN2P3, 53 Avenue
des Martyrs, F-38026 Grenoble-Cedex, France}
\author{Vincent \surname{Mathieu}}
\thanks{IISN Scientific Research Worker}
\email[E-mail: ]{vincent.mathieu@umh.ac.be} \affiliation{Groupe de
Physique Nucl\'{e}aire Th\'{e}orique, Universit\'{e} de
Mons-Hainaut, Acad\'{e}mie universitaire Wallonie-Bruxelles, Place
du Parc 20, B-7000 Mons, Belgium}

\date{\today}

\begin{abstract}
In their textbook, Suzuki and Varga [Y. Suzuki and K. Varga, {\em
Stochastic Variational Approach to Quantum-Mechanical Few-Body
Problems} (Springer, Berlin, 1998)] present the stochastic
variational method in a very exhaustive way. In this framework, the so-called
correlated gaussian bases are often employed. General formulae for the matrix
elements of various operators can be found in the textbook. However the Fourier
transform of correlated gaussians and their application to the management of a
relativistic kinetic energy operator are missing and cannot be found in the
literature. In this paper we present these interesting formulae. We give also
a derivation for new formulations concerning central potentials; the
corresponding formulae are more efficient numerically than those presented in
the textbook.
\end{abstract}

% 02.70.-c Computational techniques
% 12.39.Pn    Potential models

\maketitle

\section{Introduction}
\label{sec:intro}
There exist several different technical methods to solve the few-body problem
with accuracy: Monte Carlo calculations \cite{ham94}, Faddeev and Yakubovsky
treatments \cite{glo96}, hyperspherical formalism \cite{fab83}, expansion on
various types of orthogonal \cite{bsb93}, or non orthogonal bases \cite{suzu98}.
Each technics shows specific advantages and drawbacks.
Among others, the stochastic variational method is especially attractive. It
relies on expansion of the wave function in term of gaussian type functions.
The stochastic algorithm allows to consider very large bases with a minimum of
variational effort. The drawback of this method is the non orthogonality of
basis wave functions with the possibility of appearance of spurious states due
to overcompleteness; if this last inconvenient is overcome, using non orthogonal
bases is not really a problem. The generalized eigenvalue problem arising in
this case is well under control nowadays. The great advantage of using gaussian
type functions is the rapid convergence and, above all, the possibility to
compute the resulting matrix elements with analytical expressions most of
time.

The stochastic variational method is described in full details in the remarkable
textbook by Y. Suzuki and K. Varga \cite{suzu98}, where most important and
fundamental formulae are derived. This very complete work will be refered as SV
throughout this paper, and all subsequent references can be found in it.

However, working on other projects, we were faced to the necessity to use some
particularly important matrix elements that are not found in the exhaustive SV
textbook. In particular, the Fourier transform of the correlated gaussians and
the matrix elements of a relativistic kinetic energy operator are dramatically
missing. In this paper, we want to complete SV with important formulae that are
of crucial importance for some applications. Besides the relativistic kinetic
energy, we derive also new expressions for the matrix elements of central
potentials. These expressions are simpler and more efficient numerically than
those given in SV. We also report them here.

To achieve some unity, to precise our notations and to have a self-contained
paper, we will also present below some formulae already present in SV. In this
case we give the references where they can be found in this work. We will
derive expressions for the most general correlated gaussians (arbitrary number
of particles $N+1$, arbitrary angular momentum $L$) but we restrict ourselves
here to operators that do not mix spatial and spin degrees of freedom, so that
we are concerned only with the matrix elements for the spatial part of the wave
functions. Moreover, the non-natural parity states are very difficult to handle
in correlated bases (this is possible but the expressions are much more
involved) and, in the following, we just study natural parity (i.e spatial
parity equal to $(-1)^L$) states.

The paper is organized as follows. We first describe the systems under
consideration (intrinsic coordinates, definition of correlated gaussians
and their generating functions). Then, we give the matrix elements for
the overlap, the non-relativistic kinetic energy and a first expression for
central potentials. Except for special points, most of these formulae can be
found in SV. Lastly, the novelties concern new expressions for central
potentials, the Fourier transform of correlated gaussians and its
application to relativistic kinetic energy.

\section{The system under consideration}
\label{sec:syst}
The stochastic variational method can be applied to systems composed of more
than one particle up to around ten particles. Let us denote by $N+1$ the number
of particles ($N \geq 1)$. The particle $i$ is located at $P_i$, so that the
position vector relative to some origin $O$ is $\bm{OP}_i = \bm{r}_i$, and has a
corresponding conjugate momentum $\bm{p}_i$. Since we are concerned only with
spatial degrees of freedom, we ignore the color, isospin and spin variables.

It is interesting to introduce the center of mass coordinate $\bm{R}_{cm} =
\bm{x}_{N+1}$ and the corresponding total momentum $\bm{P} = \bm{\pi}_{N+1}$.
The intrinsic description is expressed in term of internal coordinates. There
exist different possible choices depending upon the nature of the studied
systems. Here, we are concerned only with the Jacobi coordinates.

\subsection{Jacobi coordinates}
\label{ssec:Jaccoord}

In order to simplify the notations, let us note $m_{12\ldots i} = m_1 + m_2 +
\ldots + m_i$ and $G_i$ the center of mass of the first $i$ particles:
$\bm{OG}_i = (m_1 \bm{r}_1 + m_2 \bm{r}_2 + \ldots + m_i \bm{r}_i)/m_{12\ldots i}$.
The Jacobi coordinate $\bm{x}_i$ is defined as the position of the particle $i+1$
relative to the center of mass $G_i$ of the previous particles. Explicitly
\begin{equation}
\label{eq:defcj}
\bm{x}_i = \bm{OG}_i - \bm{r}_{i+1}.
\end{equation}
There are obviously $N$ Jacobi coordinates $x_i$ ($i\leq N$)
corresponding to intrinsic coordinates and the special vector
$\bm{x}_{N+1}$ corresponds, as already mentioned, to the center of
mass coordinate for the system (this vector has a physical meaning
for particles with non vanishing masses; for the case of null mass
particles, a way to get rid of this problem is to employ a
relativistic kinetic energy, as explained in section
\ref{sec:Tr}).

It is important to express the Jacobi coordinates in terms of the original
position vectors
\begin{equation}
\label{eq:xvsr}
\bm{x}_i = \sum_{j=1}^{N+1} U_{ij}  \bm{r}_j.
\end{equation}
It is easy to calculate, from definition (\ref{eq:defcj}), the value of the
elements $U_{ij}$:
\begin{eqnarray}
\label{eq:emUij}
U_{ij} & = & \frac{m_j}{m_{12\ldots i}} \quad \textrm{if } j \leq i , \nonumber \\
U_{ii+1} & = & -1 , \\
U_{ij} & = & 0 \quad \textrm{if } j > i+1. \nonumber
\end{eqnarray}
The inversion of relation (\ref{eq:xvsr}) is often useful
\begin{equation}
\label{eq:rvsx}
\bm{r}_i = \sum_{j=1}^{N+1} (U^{-1})_{ij}  \bm{x}_j.
\end{equation}
It is a matter of simple calculation to check that the matrix elements of the
inverse matrix are given by:
\begin{eqnarray}
\label{eq:emUm1ij}
(U^{-1})_{kl} & = & \frac{m_{l+1}}{m_{12\ldots l+1}} \quad k \leq l \leq N,
 \nonumber \\
(U^{-1})_{l+1 l} & = & - \frac{m_{12\ldots l}}{m_{12\ldots l+1}} \quad
l \leq N, \\
(U^{-1})_{kl} & = & 0 \quad k > l+1, \nonumber \\
(U^{-1})_{k N+1} & = & 1 \quad \forall k. \nonumber
\end{eqnarray}

The conjugate momentum for variable $\bm{x}_i$ is denoted $\bm{\pi}_i =
-i \partial/\partial \bm{x}_i$ ($\hbar = 1)$ and must fulfill the conditions
$\left[(\bm{x}_i)_k,(\bm{\pi}_j)_l \right ] = i \delta_{ij} \delta_{kl}$.
It is easy to check that the relations corresponding to Eqs.
(\ref{eq:xvsr})-(\ref{eq:rvsx}) write

\begin{eqnarray}
\label{eq:pivsp}
\bm{\pi}_i & = &\sum_{j=1}^{N+1} (U^{-1})_{ji}  \bm{p}_j, \\
\label{eq:pvspi}
\bm{p}_i & = &\sum_{j=1}^{N+1} U_{ji}  \bm{\pi}_j.
\end{eqnarray}
With this definition, the total momentum $\bm{P} = \bm{p}_1 + \bm{p}_2
+ \ldots +  \bm{p}_{N+1}$ is just $\bm{\pi}_{N+1}$.

\subsection{Correlated gaussians}
\label{subsec:corgaus}

In the stochastic variational method, the wave function for the system is
expanded on gaussian type functions. There exist several different types of
such basis states. Here we are concerned only with the so-called ``correlated
gaussians''. The space function is expressed in terms of the Jacobi coordinates.

The simplest version is a function that is a product of separate
gaussian function for each Jacobi coordinate. In this case the
calculations are very easy, but the convergence is slow, and,
moreover, the available Hilbert space is limited, because each
pair of particles is in a S-state. A more elaborated version
considers instead the argument of the exponential as a bilinear
combination of the Jacobi coordinates: $\sum_{i,j=1}^{N} A_{ij}
\bm{x}_i \cdot \bm{x}_j$. The matrix $A$ must be symmetric ($A =
\tilde{A}$) and positive definite. Obviously, the Hilbert space is
enlarged and the convergence is accelerated.

To deal with a non vanishing total angular momentum, one must introduce
spherical harmonics somehow or other. The most elegant manner is to use a
\textbf{single} solid harmonic $\mathcal{Y}_{LM}(\bm v) = v^L Y_{LM}(\hat{v})$.
To achieve some symmetry, and also to have more variational parameters at our
disposal, the argument of the solid harmonic is the most general linear
combination of the Jacobi coordinates $\bm v = \sum_{i=1}^N u_i \bm{x}_i$.

In order to simplify the notations, let us introduce a ``super-vector''
$\tilde{\bm x} = (\bm{x}_1,\bm{x}_2,\ldots,\bm{x}_N)$ and write the
coefficients of linear combinations as a ``line (or column) matrix'', i.e
$\tilde{u} = (u_1,u_2,\ldots,u_n)$. This allows to shorten the expressions
using the usual matrix operations. For example (the presence of the symbol
$\cdot$ deals with a spatial scalar product, while the absence deals only
with a linear combination)

\begin{eqnarray}
\label{eq:notsimp}
\sum_{i,j=1}^{N} A_{ij} \bm{x}_i \cdot \bm{x}_j & = & \tilde{\bm x} \cdot A
\bm x, \nonumber \\
\sum_{i=1}^N u_i \bm{x}_i & = & \tilde{u} \bm x.
\end{eqnarray}

With those definitions, the most general correlated gaussian is
given by (note a slight difference with SV notations; their matrix
$A$ is twice ours and moreover $N$ is the number of Jacobi
coordinates while SV consider it as the number of particules)
\begin{equation}
\label{eq:corgaus}
f_{KLM}(u,A;\bm x) = \exp(- \tilde{\bm x} \cdot A \bm x)\:|\tilde{u} \bm x|^{2K}
\mathcal{Y}_{LM}(\tilde{u} \bm x).
\end{equation}
The correlated gaussian  represents the basis state in coordinate representation
$f_{KLM}(u,A;\bm x)$
 = $\left\langle \bm x | \psi_{KLM}(u,A) \right\rangle$. Thus each basis state
is described by $N(N+3)/2$ free parameters ($N(N+1)/2$ for the
matrix $A$ and $N$ for the vector $u$). This prescription
(\ref{eq:corgaus}) is only able to deal with natural parity
states. The term $|\tilde{u} \bm x|^{2K}$ is introduced for
generality and to treat with more accuracy potentials with
specific singular features. However, it complicates a lot the
resulting expressions. It is often more convenient (except when
the potential is so singular that the resulting integrals diverge)
to keep in the calculation the correlated gaussians restricted to
$K = 0$, including more basis states to compensate a slower
convergence. The even simplest version corresponds to a diagonal
$A$ matrix.

\subsection{Matrix elements and generating functions}
\label{subsec:genfunc} We are interested in the calculation of the
matrix elements for some operator $\hat{O}$ on the correlated
gaussians, namely $\langle \psi_{K'L'M'}(u',A')| \hat{O} |
\psi_{KLM}(u,A)\rangle$. The critical point in the computation of
such element is to manage the difficulties due to the presence of
the solid harmonics. An artful way to deal with them is to use the
generating functions for the correlated gaussians. Since the
technical details can be found in SV, we just recall below the
most important results and let the reader have a look on SV (in
section 6.3) for rigourous proofs.

Let us define the functions
\begin{equation}
\label{eq:genfunc}
g(\bm s,A; \bm x) = \exp(- \tilde{\bm x} \cdot A \bm x + \tilde{\bm s} \cdot
\bm x)
\end{equation}
where $\bm s$ is an arbitrary super-vector, $\tilde{\bm s} = (\bm s_1,\bm s_2,
\ldots,\bm s_N)$, and, as usual $\tilde{\bm s} \cdot \bm x$ = $\sum_{i=1}^{N}
\bm s_i \cdot \bm x_i$.

The $g$ functions are called the generating functions for the correlated
gaussians since one has
\begin{equation}
\label{eq:fvsgen}
\begin{split}
 f_{KLM}(u,A;\bm x) = &  \frac{1}{B_{KL}}\\
         \times \int d\hat{e} \:  Y_{LM}(\hat{e}) & \left(
\frac{\partial^{2K+L}}{\partial \lambda^{2K+L}} g(\lambda \bm e
u,A; \bm x)\right)_{\lambda=0, |\bm e| = 1}
\end{split}
\end{equation}
where the geometrical coefficient $B_{KL}$ is defined as
\begin{equation}
\label{eq:BKL}
B_{KL}= \frac{4 \pi (2K+L)!}{2^K \, K! \, (2K+2L+1)!!}.
\end{equation}
In Eq. (\ref{eq:fvsgen}), the super-vector $\bm s = \lambda \bm e u$ must
be understood with all its components proportional to the same three vector
$\bm e$, namely $\bm s_i = \lambda u_i \bm e$.

Using Eq. (\ref{eq:fvsgen}) in the expression of the searched matrix element
leads to
\begin{equation}\label{eq:elmatgen}
\begin{split}
\langle \psi_{K'L'M'}(u',A')| \hat{O} |\psi_{KLM}(u,A)\rangle   \\
= \frac{1}{B_{K'L'}B_{KL}}\int d\hat{e}\:d\hat{e}' Y_{LM}(\hat{e})
Y_{L'M'}^\ast(\hat{e}')
\\
\times\left( \frac{\partial^{2K'+L'+2K+L}}{\partial
{\lambda'}^{2K'+L'}
\partial \lambda^{2K+L}} \left\langle {\cal O}\right\rangle  \right)_{\lambda =
\lambda' =0,|\bm e| = |{\bm e}'| = 1}.
\end{split}
\end{equation}
with the matrix element between the generating function
\begin{equation}\label{}
\left\langle {\cal O}\right\rangle =\langle g(\lambda' {\bm e}'
u',A';\bm x) | \hat{O} | g(\lambda \bm e u,A;\bm x)\rangle
\end{equation}
The matrix element that is left for computation is now between the generating
functions, the form of which is much simpler.

In the rest of the paper, we focus our study on operators that are
scalar for spatial coordinates, so that the only non vanishing
elements are those with $L = L'$, and $M = M'$. Moreover, the
matrix elements do not depend on the magnetic quantum number $M$.

\section{Overlap}
\label{sec:overl}
The overlap between basis states $\mathcal{N}_{K'KL}$ = $\left\langle
\psi_{K'LM}(u',A') | \psi_{KLM}(u,A) \right\rangle$ is a crucial ingredient
in the equation of motion. Since the basis states are non orthogonal, there is
no reason that such an element is diagonal.

As already mentioned, it is easier to calculate first the elements for the
generating functions. The technics is always the same; we first diagonalize
the matrix $A$ and change the integration variables to the eigenvectors of
$A$. More details are presented in SV, and we will develop more deeply this
technics later on. Explicitly, one finds (see SV, Table 7.1 page 124):
\begin{eqnarray}
\nonumber
 \left\langle g(\bm{s}',A'; \bm x) |g(\bm{s},A; \bm x)
\right\rangle &=& \left( \frac{\pi^N}{\det B}\right)^{3/2}
 \exp(\frac{1}{4} \tilde{\bm v} \cdot B^{-1} \bm v)    \\\label{eq:recfgen}
 &=& \mathcal{M}_0
\end{eqnarray}
where
\begin{equation}
B = A + A'; \quad \bm v = \bm s + \bm{s}'.
\end{equation}

The overlap is obtained introducing result (\ref{eq:recfgen}) in
the general formula (\ref{eq:elmatgen}). Since the variables
$\lambda,\lambda',\hat{e}, \hat{e}'$ appear only in $\bm v$, the
term depending on $\det B$ factorizes out the integral. The
exponential is then expanded as a series in $\lambda,\lambda', \bm
e \cdot \bm e'$. The derivation and integration are performed
without difficulty. The final result is (SV, (A.6) page 248)

\begin{eqnarray}
\label{eq:recgen} \mathcal{N}_{K'KL}=\frac{(2K'+L)! \:
(2K+L)!}{B_{K'L}B_{KL}}\left( \frac{\pi^N}{\det B}\right)^{3/2}
\\\nonumber \times\!\!\! \sum_{k=0}^{\min (K,K')}
B_{kL}
\frac{q^{K-k}}{(K-k)!}\frac{q'^{K'-k}}{(K'-k)!}\frac{\rho^{2k+L}}{(2k+L)!}
\end{eqnarray}
with the following complementary definitions
\begin{eqnarray}
\nonumber
  q &=& \frac{1}{4} \tilde{u}B^{-1}u; \quad q'=
\frac{1}{4} \tilde{u}'B^{-1}u'; \\
  \rho &=&\frac{1}{2} \tilde{u}'B^{-1}u = \frac{1}{2}
\tilde{u}B^{-1}u'.\label{eq:defqqprho}
\end{eqnarray}
For the important peculiar case $K' = K = 0$, this formula
simplifies a lot and we are left with (SV (A.7) page 249):

\begin{equation}
\label{eq:recpart}
\mathcal{N}_{00L}= \mathcal{N}_L = \frac{(2L+1)!!}{4 \pi}\left(
\frac{\pi^N}{\det B}\right)^{3/2} \rho^L.
\end{equation}

\section{Non relativistic kinetic energy}
\label{sec:Tnr}
It is well known that the use of Jacobi coordinates allows to share the total
kinetic energy operator $T = \sum_{i=1}^{N+1} \bm p_i^2/(2 m_i)$ into the
kinetic energy of the system as a bulk $T_{cm} = \bm P^2/(2 M)$ ($M =
m_{12\ldots N+1}$) and an intrinsic kinetic energy operator $T_{NR}$
depending only on intrinsic momenta.

A simple calculation leads to the explicit expression
\begin{eqnarray}
 \nonumber
 T_{NR}   &=&  T - T_{cm}\\
   &=&\frac{1}{2} \sum_{i,j=1}^N \Lambda_{ij} \bm
\pi_i \cdot \bm \pi_j = \frac{1}{2} \tilde{\bm \pi} \cdot \Lambda
\bm \pi\label{eq:Tnr1}
\end{eqnarray}
where the super-vector $\tilde{\bm \pi}=(\bm \pi_1,\bm
\pi_2,\ldots,\bm \pi_N)$ is introduced. The matrix $\Lambda$ is
symmetric and its general element writes
\begin{equation}
\label{eq:lambdaij}
\Lambda_{ij} = \sum_{k=1}^{N+1} U_{ik} U_{jk} \frac{1}{m_k}.
\end{equation}

The matrix elements of $T_{NR}$ on correlated gaussians rely again on the
expression of the elements on the generating functions. This calculation
is a bit more complicated, but one can show that (see SV, Table 7.1 page 124):
\begin{equation}
\begin{split}
\langle g(\bm{s}',A'; \bm x) | \tilde{\bm \pi} \cdot \Lambda
\bm \pi| & g(\bm{s},A; \bm x) \rangle = \\
& \mathcal{M}_0 \left[6 \textrm{Tr}(AB^{-1}A' \Lambda) -
\tilde{\bm y} \cdot \Lambda \bm y \right]\label{eq:Tnrfgen}
\end{split}
\end{equation}
where $\mathcal{M}_0$ is precisely the overlap expression
\eqref{eq:recfgen}, Tr($X$) means the trace of the $X$ matrix, and
the super-vector $\bm y$ is given by
\begin{equation}
\label{eq:defsvy}
\bm y = A'B^{-1} \bm s - A B^{-1} \bm s'.
\end{equation}

It remains to insert result (\ref{eq:Tnrfgen}) into Eq. (\ref{eq:elmatgen}).
The variables $\lambda,\lambda', \bm e \cdot \bm e'$ appear now still in the
exponential part of $\mathcal{M}_0$ but also  in $\tilde{\bm y} \cdot \Lambda
\bm y $ while the trace is independent of these. Fortunately, the new contribution
is just a polynomial in term of these variables. Thus, developping the exponential
as a series leads to a total contribution which is polynomial and which can be
treated exactly in the same way as the overlap.

The final result is (see SV, (A.10) page 250)
\begin{widetext}
\begin{eqnarray}
\label{eq:Tnrgen} \left\langle \psi_{K'LM}(u',A') | \tilde{\bm\pi}
\cdot \Lambda \bm \pi | \psi_{KLM}(u,A) \right\rangle
=\frac{(2K'+L)! \: (2K+L)!}{B_{K'L}B_{KL}} \left(\frac{\pi^N}{\det
B}\right)^{3/2} \nonumber\\
\times \sum_{k=0}^{\min (K,K')} B_{kL} \left[ Rqq'\rho +
P(K-k)q'\rho + P'(K'-k)q\rho + Q(2k+L)qq'\right]
\frac{q^{K-k-1}}{(K-k)!} \frac{q'^{K'-k-1}}{(K'-k)!}
\frac{\rho^{2k+L-1}}{(2k+L)!}
\end{eqnarray}
\end{widetext}
with the numbers $P,P',Q,R$ defined by
\begin{eqnarray}
\label{eq:PPQR}
P=-\tilde{u}B^{-1}A'\Lambda A'B^{-1}u&;& \quad
P'=-\tilde{u'}B^{-1}A\Lambda AB^{-1}u'; \nonumber \\
Q = 2\tilde{u'}B^{-1}A\Lambda A'B^{-1}u&;& \quad
R = 6 \textrm{Tr}(AB^{-1}A' \Lambda).
\end{eqnarray}
Again, the formula for the special case $K = K' = 0$ is much
simpler
\begin{equation}
\label{eq:Tnrpart}
\left\langle \psi_{0LM}(u',A') | \tilde{\bm \pi} \cdot \Lambda \bm \pi |
\psi_{0LM}(u,A) \right\rangle = \mathcal{N}_L \left(R + L \frac{Q}{\rho}\right).
\end{equation}
The pleasant feature is the factorization of the overlap matrix.

Clearly the matrix $\tilde{U}$, transpose of $U$ defined in Eq. (\ref{eq:emUij}),
is different from $U^{-1}$ defined in Eq. (\ref{eq:emUm1ij}). Thus $U$ is not an
orthogonal matrix. Nevertheless, one has the property $\det U = 1$. This matrix
has also another very interesting property.

Let us define the reduced mass $\mu_i$ for particle $i$ by
\begin{equation}
\label{eq:massred}
\frac{1}{\mu_i} = \frac{1}{m_{12\ldots i}} + \frac{1}{m_{i+1}}.
\end{equation}
It can be shown that the definition (\ref{eq:xvsr}) allows to
write the non-relativistic kinetic energy operator in a form
without mixed terms, namely
\begin{equation}
\label{eq:Tnrform}
T_{NR}= \frac{1}{2} \tilde{\bm \pi} \cdot \Lambda \bm \pi =
\sum_{i=1}^N \frac{\bm \pi_i^2}{2 \mu_i}.
\end{equation}
This means that, in the matrix elements (\ref{eq:Tnrgen})-(\ref{eq:Tnrpart}),
one can restrict the calculations to a diagonal matrix $\Lambda$ since, in
this case, $\Lambda_{ij} = (1/\mu_i) \delta_{ij}$.

\section{Central potentials: a first expression}
\label{sec:centpot1} In the most interesting cases, the potentials
appearing in the few-body problem are either one-body potentials
or two-body potentials. In the first situation, the potential has
a form $V=\sum_i V_i(\bm r_i - R_{cm})$ while the second situation
the potential has a form $V= \sum_{i<j} V_{ij} (\bm r_i-\bm r_j)$.
This is the consequence of translational invariance. In both
cases, the argument of any potential is always a linear
combination of Jacobi coordinates, so that the most general form
of the potential is a sum of terms like $V(\tilde{w} \bm x)$.

In this case, one can write generally
\begin{equation}
\label{eq:elpotgen} \left\langle \psi|V(\tilde{w} \bm x) |\psi
\right\rangle = \int V(\bm r) \left\langle
\psi|\delta^3(\tilde{w}\bm x - \bm r)| \psi \right\rangle \: d\bm
r .
\end{equation}
It is necessary to calculate the matrix elements of
$\delta(\tilde{w} \bm x - \bm r)$ on the generating functions;
these are given in SV Table 7.1 page 124.

However, we are interested here in central potentials that are also invariant
under rotations; the argument of the potential must be now $|\tilde{w} \bm x|$.
In this case, the equivalent expression of (\ref{eq:elpotgen}) needs a single
integral
\begin{equation}
\label{eq:elpotcent} \left\langle \psi|V(|\tilde{w} \bm x|)|\psi
\right\rangle = \int V(r) \left\langle \psi |\,\delta(|\tilde{w}
\bm x| - r) \, | \psi\right\rangle \: dr.
\end{equation}
It is necessary to calculate the matrix elements of
$\delta(|\tilde{w} \bm x| - r)$ on the generating functions. They
are not given in SV. It is possible to get them either directly by
elementary integrations, or by the developing $\delta^3(\tilde{w}
\bm x - \bm r)$ in spherical coordinates and integrating on
angular variables. We find
\begin{equation}\label{eq:deltscalfgen}
\begin{split}
\left\langle g(\bm{s}',A'; \bm x)|\delta(|\tilde{w} \bm x| - r) |
g(\bm{s},A; \bm x)
\right\rangle  & = \frac{4 \mathcal{M}_0}{\sqrt{\pi}}\\
 \times\frac{r^2}{(\tilde{w}B^{-1}w)^{3/2}} i_0 \left(\frac{r
|\tilde{w}B^{-1} \bm v|}{\tilde{w}B^{-1}w}\right)  \exp & \left(-
\frac{r^2 + (\tilde{w}B^{-1} \bm v)^2/4}{\tilde{w}B^{-1}w}\right)
\end{split}
\end{equation}
where $i_0(z) = \sinh(z)/z$ is a modified spherical Bessel function.

One must implement the element (\ref{eq:deltscalfgen}) into Eq. (\ref{eq:elmatgen}).
For further convenience, we give here more information concerning this computation.
Let us note $c = 2/(\tilde{w}B^{-1}w)$. First, it is possible to remove the
term $\exp(-c r^2/2)$ and all terms depending on $c$ only out of the derivation
and, even, out of the integral. The terms containing the variables $\lambda,
\lambda',\bm e \cdot \bm e'$ are present in the exponential part of
$\mathcal{M}_0$ which writes explicitly
\begin{equation}
\label{eq:expqq}
\mathcal{M}_0 \propto \exp(q \lambda^2 + q' {\lambda'}^2+
\rho \lambda \lambda' \bm e \cdot \bm e').
\end{equation}
These terms also appears in the variable $\bm z = (\tilde{w}B^{-1}
\bm v)/2$. Let us introduce the numbers $\gamma = c
(\tilde{w}B^{-1}u)/2$, $\gamma' = c (\tilde{w}B^{-1}u')/2$. The
variable $\bm z$ now reduces to $c \bm z = \gamma \lambda \bm e +
\gamma' \lambda' \bm e' $. Thus, the term $i_0(cr|\bm z|) \exp(-c
\bm z^2/2)$ must also be included in the derivation term.

Using the fact that $\exp(-s^2+2sx)$ is the generating function for Hermite
polynomials $H_n(x)$, it is possible to prove the interesting relation
\begin{equation}
\label{eq:genpolher} i_0(Bz) \exp(-Az^2) = \frac{\sqrt{A}}{B}
\sum_{n=0}^\infty \frac{(Az^2)^n} {(2n+1)!} H_{2n+1}(B/2\sqrt{A}).
\end{equation}

The technics presented in SV consists in employing this formula to evaluate
$i_0(cr|\bm z|) \exp(-c \bm z^2/2)$. Forgetting terms depending on $c$ only,
there appears a term depending on $r$: $H_{2n+1}(r\sqrt{c/2})/r$ and a term
\begin{equation}
\label{eq:lienzee}
(c^2 \bm z^2)^n = (\gamma^2 \lambda^2 + {\gamma'}^2 {\lambda'}^2 + 2 \gamma
\gamma' \lambda \lambda' \bm e \cdot \bm e')^n
\end{equation}
which must be maintained in
the derivation and which is expanded as a series. It is gathered with the
series coming from exponential (\ref{eq:expqq}). After this grouping, the
general monome of this series has the form $\lambda^s {\lambda'}^{s'} (\bm e
\cdot \bm e')^t$. The derivation $\partial^{2K+L}/\partial \lambda^{2K+L}$
being taken for $\lambda = 0$ leads to a term $(2K+L)! \, \delta_{2K+L,s}$;
one has also a term $(2K'+L)! \, \delta_{2K'+L,s'}$ coming from the derivation
with respect to $\lambda'$. Lastly, the term $(\bm e \cdot \bm e')^t$ is
expanded as (do not forget that $|\bm e|$ = 1 = $|\bm e'|$)

\begin{equation}
\label{eq:eepdel}
(\bm e \cdot \bm e')^t = 4 \pi \sum_{lm} \frac{t!}{(t-l)!!(t+l+1)!!}
Y_{lm}^\ast(\hat{e}) Y_{lm}(\hat{e}')
\end{equation}
and the integration over $\hat{e}$ and $\hat{e}'$ leads to a term $B_{kL} \,
\delta_{2k+L,t}$.

After this tedious but straightforward cooking, and using
expression (\ref{eq:elpotcent}), we get the final result (SV,
(A.128) page 282):
\begin{equation}
\label{eq:Vcentgen}
\begin{split}
\langle &\psi_{K'LM}(u',A') | V(|\tilde{w} \bm x|) |
\psi_{KLM}(u,A) \rangle =\\& \left(\frac{\pi^N}{\det
B}\right)^{3/2}\frac{(2K'+L)! \: (2K+L)!}{B_{K'L}B_{KL}}
\sum_{n=0}^{K+K'+L} \frac{J(n,c)}{c^n}\\
&\times   \sum_{k=0}^{\min (K,K')}  B_{kL}
F^n_{K-k,K'-k,2k+L}(q,q',\rho,\gamma,\gamma')
\end{split}
\end{equation}
In this formula, the dynamical quantities $q,q',\rho$ have be
defined in Eq. (\ref{eq:defqqprho}) for the overlap, while the new
ones $c,\gamma,\gamma'$ are specific to the potential
\begin{equation}
\label{eq:cgamgam}
c=\frac{2}{\tilde{w}B^{-1}w}; \quad \gamma = \frac{\tilde{w}B^{-1}u}
{\tilde{w}B^{-1}w};\quad \gamma' = \frac{\tilde{w}B^{-1}u'}{\tilde{w}B^{-1}w}.
\end{equation}
The form of the potential appears explicitly through the following integral
\begin{equation}
\label{eq:Jnc} J(n,c)=\!\!\!\frac{1}{\sqrt{\pi}(2n+1)!}
\int\limits_0^\infty \!\!V(x \sqrt{2/c}) e^{-x^2} H_1(x)
H_{2n+1}(x) \: dx.
\end{equation}
Lastly, the geometrical function $F$ is defined as
\begin{equation}\label{eq:fFVarg}
\begin{split}
F^n_{p,p',l}(q,q',\rho,\gamma,\gamma')=n! \sum_{m,m'}
\frac{q^{p-m}}{(p-m)!} \frac{{q'}^{p'-m'}} {(p'-m')!}\\ \times
\frac{\rho^{l-n+m+m'}}{(l-n+m+m')!}\frac{\gamma^{n+m-m'} {\gamma
'}^{n-m+m'}}{2^{m+m'}\,m!\,m'!(n-m-m')!}.
\end{split}
\end{equation}
We tried to express this function in term of hypergeometric
functions, but without success (when no explicit bounds are
indicated, the indices in the summations run on values that do not
give integer negative values for the factorials).

The expression (\ref{eq:Vcentgen}) has the nice property to allow easy
checks. The bra-ket symmetry is trivial. For such a symmetry, $K \leftrightarrow
K'$, $A \leftrightarrow A'$, $u \leftrightarrow u'$, $q \leftrightarrow q'$,
$\gamma \leftrightarrow \gamma'$, while $B$, $\rho$ and $c$ remain unchanged.
The symmetry follows from the property $F^n_{p,p',l}(q,q',\rho,\gamma,\gamma')$
= $F^n_{p',p,l}(q',q,\rho,\gamma',\gamma)$. It is also easy to recover the
overlap expression if we set $V(r) = 1$. In this case, due to the orthogonality
of Hermite polynomials, one has $n = 0$, $J(n,c)/c^n = 1$, $F^0_{p,p',l}$ =
$q^p {q'}^{p'} \rho^l/[p! \, p'! \, l!]$ and the overlap expression follows
immediately.

As always, there is a great simplification in the special case $K
= K' = 0$. Indeed, one has the simple expression (SV, (A.130) page
282)
\begin{equation}
\label{eq:Vcentpart} \left\langle \psi_{0LM} | V(|\tilde{w} \bm
x|) | \psi_{0LM} \right\rangle = \mathcal{N}_L \: L! \sum_{n=0}^L
\frac{J(n,c)}{(L-n)!} \left(\frac{\gamma \gamma'}{\rho
c}\right)^n.
\end{equation}
Here again, the overlap factorizes.

A small drawback in the formulae (\ref{eq:Vcentgen}) and
(\ref{eq:Vcentpart}) is the presence of the Hermite polynomial
$H_{2n+1}(x)$. One way to deal with it is the use of recursion
formulae. However, the accuracy decreases with increasing $n$;
moreover, closed expressions for $J(n,c)$ do not exist for various
forms of potential $V(r)$. There is a way to get rid of these two
difficulties.

We define new integrals:
\begin{equation}
\label{eq:intFV}
\mathcal{F}_V(k,A) = \int_0^\infty V(u) u^k e^{-Au^2} \: du.
\end{equation}
Indeed, a lot of closed expressions exist for various forms of potentials $V(u)$.
Expanding the Hermite polynomials as a series, and rearranging the summations leads
to a new expression for the matrix elements in terms of $\mathcal{F}_V$ functions.
This form is absent in SV.
\begin{widetext}
\begin{equation}
\begin{split}
\label{eq:Vcentgen2} \left\langle \psi_{K'LM}(u',A') |
V(|\tilde{w} \bm x|) | \psi_{KLM}(u,A) \right\rangle =
\frac{(2K'+L)! \: (2K+L)!}{\sqrt{2\pi} 2^{K+K'-1} B_{K'L}B_{KL}}
  \left(\frac{c \pi^N}{\det B}\right)^{3/2} \frac{\gamma^{2K+L}
{\gamma'}^{2K'+L}}{(-c)^{K+K'+L}} \\ \times\sum_{n=0}^{K+K'+L}
\frac{(-2c)^n \mathcal{F}_V(2n+2,c/2)}{(2n+1)!} \sum_{k=0}^{\min
(K,K')} 4^k B_{kL} H^{K,K',L}_{n,k} \left( \frac{2q \gamma'}{\rho
\gamma},\frac{2q' \gamma}{\rho \gamma'} , \frac{\rho c}{\gamma
\gamma'} \right)
\end{split}
\end{equation}
\end{widetext}
with function $H$ defined by
\begin{equation}
\label{eq:fHnk} \begin{split} &H^{K,K',L}_{n,k}(x,x',y)=
\\ &\sum_{r=0}^{K+K'+L-n} (-1)^r \frac{(K+K'+L-r)! \: y^r}
{(K+K'+L-n-r)!} G^{K,K',L}_{n,k,r}(x,x')
\end{split}
\end{equation}
and
\begin{equation}
\label{eq:fGnkr}
\begin{split}
G^{K,K',L}_{n,k,r}(x,x')& = \sum_{s=0}^{K-k} \sum_{s'=0}^{K'-k}
\frac{x^s {x'}^{s'}}{s! (K-k-s)! s'! (K'-k-s')! } \\
&\times\frac{1}{(r-s-s')!(2k+L+s+s'-r)!}.
\end{split}
\end{equation}
Due to the property $G^{K,K',L}_{n,k,r}(x,x')=G^{K',K,L}_{n,k,r}(x',x)$ the
hermiticity of the element is transparent.

For the special case $K = K' = 0$, this formula reduces to
\begin{equation}
\label{eq:Vcentpart2}
\left\langle \psi_{0LM}(u',A') | V(|\tilde{w} \bm x|) |
\psi_{0LM}(u,A) \right\rangle = \mathcal{N}_L \: 2c \sqrt{\frac{c}{2 \pi}} \: L!
\nonumber
\end{equation}
\begin{equation}
\times \sum_{n=0}^L \frac{2^n \mathcal{F}_V(2n+2,c/2)}{(2n+1)!(L-n)!}
\left(\frac{\gamma \gamma'}{\rho}\right)^n \left(1 - \frac{\gamma \gamma'}
{\rho c}\right)^{L-n}.
\end{equation}
This formula needs essentially the same numerical effort than
expression (\ref{eq:Vcentpart}) but it is written in terms of a
much more convenient integral.

\section{Central potentials: a new expression}
\label{sec:centpot2}
In this section we present new formulae for the matrix elements of a central
potential; the corresponding expressions are more efficient for a numerical
treatment.

Let us remark that the relationship (\ref{eq:lienzee}) gives a link between
$\bm e \cdot \bm e'$ and the variable $\bm z$. This opportunity allows to
put the exponential (\ref{eq:expqq}) under the form
\begin{equation}
\label{eq:expqq2}
\exp(\bar{q} \lambda^2 + \bar{q}\,' {\lambda '}^2 + \rho c^2 \bm z^2/
(2 \gamma \gamma') )
\end{equation}
with the new variables $\bar{q} = q - \rho \gamma /(2 \gamma')$
and \makebox{$\bar{q}\,' = q' - \rho \gamma' /(2 \gamma)$}.

One gathers the part of exponential depending upon $\bm z$ with the other
exponential depending upon $\bm z$ into a single exponential which takes the
form $\exp(-\bar{c} \bm z^2/2)$ with the definition $\bar{c} = c(1 - \rho c/
(\gamma \gamma'))$. We then treat the term $i_0(cr|\bm z|) \exp(-\bar{c}
\bm z^2/2)$ with the already mentioned relation (\ref{eq:genpolher}). The
rest of the derivation is essentially similar to that of section
\ref{sec:centpot1}. The trick of gathering two exponentials into a single one
allows to gain one expansion into a series; the price to pay is the use
of renormalized dynamical quantities.

The final formula, which is absent in SV, looks like
\begin{equation}
\label{eq:Vcentgen3}
\begin{split}
&\left\langle \psi_{K'LM}(u',A') | V(|\tilde{w} \bm x|) |
\psi_{KLM}(u,A) \right\rangle = \\
&\frac{(2K'+L)! \: (2K+L)!}{B_{K'L}B_{KL}} \left(\frac{\alpha
\pi^N}{\det B}\right)^{\frac{3}{2}} \\ & \times
\sum_{n=0}^{K+K'+L}
\left(\frac{\alpha}{2c}\right)^n J(n,\alpha,c) \\
&\times \sum_{k=0}^{\min (K,K')}  \frac{2^{2k+L}}{(2k+L)!} B_{kL}
F^{K,K',L}_{n,k}(\bar{q},\bar{q}\,',\gamma,\gamma').
\end{split}
\end{equation}
In this formula, the dynamical quantities $c,\gamma,\gamma'$ are identical
to those defined in Eq. (\ref{eq:cgamgam}) while new quantities are necessary
\begin{equation}
\label{eq:quantbar}
\bar{q} = q-\frac{\rho \gamma}{2 \gamma'}; \quad \bar{q}\,' = q'-
\frac{\rho \gamma'}{2 \gamma}; \quad \alpha = 1 - \frac{\rho c}{\gamma \gamma'}.
\end{equation}
The form of the potential appears explicitly through an integral of new type
\begin{equation}
\label{eq:Jnac}
\begin{split} J(n,&\alpha,c)=\frac{1}{\sqrt{\pi}(2n+1)!}\\ &
\times \int\limits_0^\infty V(x \sqrt{2\alpha/c}) e^{- \alpha x^2}
H_1(x) H_{2n+1}(x) \: dx. \end{split}
\end{equation}
Lastly, the geometrical function $F$ is defined by
\begin{equation}
\label{eq:fFmoi}
\begin{split}
 &F^{K,K',L}_{n,k}(x,x',y,y')=n! \sum_{m=\max(k+L,n-K')}^{\min(n-k,K+L)}
 \frac{x^{K+L-m}}{(K+L-m)!}\\
 &\times \frac{{x'}^{K'-n+m}}{(K'-n+m)!} \frac{y^{2m-L}{y'}^{2(n-m)+L}}
{(m-k-L)!(n-k-m)!}.
\end{split}
\end{equation}
Comparison between equivalent forms (\ref{eq:Vcentgen}), given
in SV, and (\ref{eq:Vcentgen3}), novelty of this paper, warrants some
comments.
\begin{itemize}
    \item They both have very similar aspect. The new form requires the calculations
of 3 new quantities $\bar{q},\bar{q}\,',\alpha$ and an integral $J(n,\alpha,c)$
instead of $J(n,c)$. But in doing this, the numerical effort is essentially the
same. In contrast, the rest of the calculation is much more efficient under the
new version; indeed the $F$ function depends on 4 variables instead of 5, it
requires a single summation instead of a double and the general monome needs
one factorial less.
\item The equivalence between both can be proved using the following very
interesting relationship
\begin{equation}
\label{eq:relHerm} \begin{split} & \frac{1}{a(a^2 - 1)^L }
\int\limits_0^\infty f(y) e^{-y^2} H_1(y) H_{2L+1}(ay) \: dy  =\\
& \sum_{n=0}^L \left(\frac{a^2}{a^2-1}\right)^n
\frac{(2L+1)!}{(L-n)!(2n+1)!} \\ & \times
 \int\limits_0^\infty f(y) e^{-y^2} H_1(y) H_{2n+1}(y) \: dy
 \end{split}
\end{equation}
valid for any function $f(y)$. \item The symmetry properties are
more transparent in the SV version, but they can be proved as well
in the new version. Once the equivalence between both is shown
this point is of minor importance.
\end{itemize}

The new version is especially interesting in the peculiar case $K = K' = 0$ since
the result writes
\begin{equation}
\label{{eq:Vcentpart2}} \left\langle \psi_{0LM}| V(|\tilde{w} \bm
x|) | \psi_{0LM} \right\rangle = \mathcal{N}_L \: L!
\left(\frac{\alpha}{1-\alpha} \right)^L \alpha^{3/2}
J(L,\alpha,c).
\end{equation}
In this case, there is no summation at all!

To finish this part, let us derive the general formula expressed
in term of the $\mathcal{F}$ integral instead of the $J$ integral.
The calculation relies essentially on the technics presented in
section \ref{sec:centpot1}. The result is

\begin{equation}
\label{eq:Vcentgen4} \begin{split} &\left\langle
\psi_{K'LM}(u',A') V(|\tilde{w} \bm x|) | \psi_{KLM}(u,A)
\right\rangle =\\& \frac{4 (2K'+L)! \: (2K+L)!}{\sqrt{\pi}
B_{K'L}B_{KL}} \left(\frac{c \pi^N}{2 \det B}\right)^{3/2} \\ &
\times \sum_{n=0}^{K+K'+L} \frac{\mathcal{F}_V(2n+2,c/2)}{(2n+1)!} \\
&\times \sum_{k=0}^{\min (K,K')}\frac{2^{2k+L}}{(2k+L)!} B_{kL}
G^{K,K',L}_{n,k}
 (\bar{q},\bar{q}\,',\gamma,\gamma',c)
\end{split}
\end{equation}
with function $G$ defined by
\begin{eqnarray}
\label{eq:fHnk}
 & & G^{K,K',L}_{n,k}(\bar{q},\bar{q}\,',\gamma,\gamma',c)= \\
 & &\sum_{r=k+L}^{K+L} \frac{\bar{q}^{K+L-r}\bar{q}\,'^{K'+r-n} \gamma^{2r-L}
{\gamma '}^{2(n-r)+L}} {(K+L-r)!(r-k-L)!} M^{K'}_{n,k,r}(- \frac{\alpha
{\gamma '}^2}{2 c \bar{q}\,'}) \nonumber
\end{eqnarray}
and with
\begin{equation}
\label{eq:fGnkr} M^{K'}_{n,k,r}(z) =\!\!\!\!\!\!\!\!\!\!\!\!\!\!\!
\sum_{s=\max(0,k+r-n)}^{K'+r-n} \!\!\!\!\!
\frac{(s+n)!}{(K'+r-s-n)!(s+n-k-r)!} \frac{z^s}{s!}.
\end{equation}

This formulation is also more efficient numerically than (\ref{eq:Vcentgen2})
for the same reasons as those explained above.

Application of this formula to the special case $K=K'=0$ leads exactly to the
formula (\ref{eq:Vcentpart2}) got before and thus does not bring any novelty.

\section{Fourier transform of correlated gaussians}
\label{sec:Fourtrans}
For a number of applications, it is useful to have the Fourier transform (FT)
of correlated gaussians. This point is missing in SV and, to our knowledge,
such an expression seems absent in the literature. In this section, we want to
derive the corresponding relation.

The basic ingredient for this calculation is the FT of a correlated gaussian
limited to one variable, namely the value of the integral
\begin{equation}
(2 \pi)^{-3/2}
\int e^{-i \bm q \cdot \bm r} e^{- \alpha r^2} r^{2K} \mathcal{Y}_{LM}(\bm r)
\: d \bm r.
\end{equation}

The first thing to do is to use the traditional development of a plane wave in
terms of spherical harmonics $e^{-i \bm q \cdot \bm r}$ = $4 \pi \sum_{l=0}^
{\infty} (-i)^l j_l(qr) \sum_{m = -l}^{+l} Y_{lm}(\hat{q})
Y_{lm}^\ast(\hat{r})$ where $j_l(z) = \sqrt{\pi/(2z)} J_{l+1/2}(z)$ is the
spherical Bessel function. Integration over angular $\hat{r}$ variables
leads to a $\delta_{lm,LM}$ factor which restricts the infinite summation to
the single term $l=L,m=M$. Apart from some constant factors that can be gathered
outside the integral, we are left with a radial integral, which, after a
trivial change of variable, can be put under the form $\int_0^\infty
e^{-x^2} x^{2K+L+3/2} J_{L+1/2}(qx/\sqrt{\alpha})$. Fortunately, this
integral is analytical and can be expressed in terms of generalized Laguerre
polynomial $L_n^a$. This miraculous formula can be found in Ref. \cite{grad94}
(formula (6.631) n°\, 10,page 738); it writes (it can be obtained using the
generating function of Laguerre polynomials expressed in term of Bessel
functions
\cite{abra65})
\begin{equation}
\label{eq:intlag}
\int_0^\infty e^{-x^2} x^{2n+\mu+1} J_\mu(2x\sqrt{z}) \: dx = \frac{n!}{2}
e^{-z}z^{\mu/2} L_n^{\mu}(z).
\end{equation}
In this formula $\mu$ is any real number $> -1$ and $n$ is a positive or null
integer. The case $n=0$ is simpler due to the property $L_0^a(z)=1, \forall z$.
We thus get the searched integral

\begin{equation}
\label{eq:TFcorsimp}
\begin{split}&
\frac{1}{(2 \pi)^{3/2}}
 \int e^{-i \bm q \cdot \bm r} e^{- \alpha r^2} r^{2K} \mathcal{Y}_{LM}(\bm r)
\: d \bm r =  \\  & \frac{K!}{\alpha^K} L_K^{L+1/2}
\left(\frac{q^2} {4 \alpha}\right)\frac{(-i)^L}{(2
\alpha)^{L+3/2}}e^{- q^2/(4 \alpha)} \mathcal{Y}_{LM}(\bm q).
\end{split}
\end{equation}
To get the formula corresponding to $K=0$, it is sufficient to replace in
Eq. (\ref{eq:TFcorsimp}) the terms depending on $K$ by 1, at the head of the
right hand side expression.

To derive the general FT, we begin, for pedagogical reasons, with the
special case $K = 0$. Let us note
\begin{eqnarray}
\label{eq:defTFK0}
 & & h_{0LM}(u,A;\bm \pi) =
(2 \pi)^{-3N/2} \int e^{-i \tilde{\bm \pi} \cdot \bm x} f_{0LM}(u,A;\bm x)
 \: d \bm x  \nonumber \\
 & & =(2 \pi)^{-3N/2} \int e^{- \tilde{\bm x} \cdot A \bm x -i \tilde{\bm \pi}
\cdot \bm x} \mathcal{Y}_{LM}(\tilde{u} \bm x) \: d \bm x = I.
\end{eqnarray}

The matrix $A$ is symmetric and thus can be diagonalized with an orthogonal
matrix $T$ : $A = TD\tilde{T}$. $A$ being positive definite, all the
eigenvalues of the diagonal matrix $D$ are positive, and there is no problem
to build the square root $D^{1/2}$ of this matrix. Instead of the original
$\bm x$ variables, it is wily to work with the new variables $\bm z = D^{1/2}
\tilde{T} \bm x$. At this stage, one has
\begin{equation}
\label{eq:FTwz}
\begin{split}
I &=(2 \pi)^{-3N/2} (\det A)^{- 3/2} \\ &\times\int d \bm z \:
\exp(- \tilde{\bm z} \cdot \bm z - i \tilde{\bm \pi} \cdot
TD^{-1/2} \bm z) \mathcal{Y}_{LM}(\tilde{u}TD^{-1/2} \bm z).
\end{split}
\end{equation}

In a second step, we change again from $\bm z$ to new $\bm Z$ variables with
the help of an orthogonal matrix $U$ : $\bm Z = U \bm z$. The Jacobian of
the transformation is unity, and, moreover, $\tilde{\bm z} \cdot \bm z =
\tilde{\bm Z} \cdot \bm Z$. There is a large freedom in the choice of the $U$
matrix. One can take this opportunity making a choice such that the argument
of the solid harmonic is proportional to one of the new variables, let us say
$\bm Z_1$. Now, the integral takes the form:
\begin{equation}
\label{eq:FTwz2}
\begin{split}
I =&(2 \pi)^{-3N/2} \alpha^{-L} (\det A)^{- 3/2}\\&\times \int d
\bm Z \: \exp(- \tilde{\bm Z} \cdot \bm Z - i \tilde{\bm V} \cdot
\bm Z) \mathcal{Y}_{LM}(\bm Z_1).
\end{split}
\end{equation}
with $\alpha = (\tilde{u}A^{-1}u)^{-1/2}$ and $\bm V = U D^{-1/2}\tilde{T}
\bm \pi$.

Since the integral is separable for each variable, the rest of the proof is
quite easy. $N-1$ integrals, corresponding to $\bm Z_k, k > 1$ are equal to
\begin{equation}
\label{eq:FTwzk}
(2 \pi)^{-3/2} \int d \bm Z_k \exp(- \bm Z_k^2 - i \bm V_k \cdot \bm Z_k) =
2^{-3/2} \exp(-\bm V_k^2/4)
\end{equation}
while the integral relative to $\bm Z_1$ can be calculated using expression
(\ref{eq:TFcorsimp}) to get
\begin{eqnarray}
\label{eq:FTwzk}
 & &(2 \pi)^{-3/2} \int d \bm Z_1 \exp(- \bm Z_1^2 - i \bm V_1 \cdot \bm Z_1)
\mathcal{Y}_{LM}(\bm Z_1) = \nonumber \\
 & &(- i)^L 2^{-L-3/2} \exp(-\bm V_1^2/4) \mathcal{Y}_{LM}(\bm V_1).
\end{eqnarray}

Grouping all the exponentials into a single one gives an exponential with
the argument $- (\bm V_1^2 + \bm V_2^2 + \ldots \bm V_N^2)/4$ = $- (\tilde{\bm V}
\cdot \bm V)/4$ = $- (\tilde{\bm \pi} \cdot A^{-1} \bm \pi)/4$. Thus, the final
result for the FT looks like
\begin{equation}
\label{eq:FTcasK=0} h_{0LM}(u,A;\bm \pi) =
\frac{(-i)^Le^{-\frac{1}{4} \tilde{\bm \pi} \cdot A^{-1} \bm
\pi}}{2^{L+3N/2} (\det A)^{3/2}}
\mathcal{Y}_{LM}(\tilde{u}A^{-1}\bm \pi).
\end{equation}

It is remarkable that, \textbf{in this special case} $K=0$, \textbf{the FT of a
correlated gaussian is proportional to a correlated gaussian} since we
have the property
\begin{equation}
\label{eq:FTcasK=0cg} h_{0LM}(u,A;\bm \pi) =
\frac{(-i)^L}{2^{L+3N/2} (\det A)^{3/2}}
f_{0LM}(A^{-1}u,\frac{A^{-1}}{4};\bm \pi).
\end{equation}
With this relationship, it is instructive to check the
conservation of the norm, namely $\mathcal{N}_L$ = $\left\langle
f_{0LM}| f_{0LM} \right\rangle$ = $\left\langle h_{0LM} | h_{0LM}
\right\rangle$.

We are interested now in the expression of the FT for the most general correlated
gaussian
\begin{eqnarray}
\label{eq:defTFgen}
 & & h_{KLM}(u,A;\bm \pi) = \left\langle \bm \pi | \psi_{KLM}(u,A)
\right\rangle = \nonumber \\
 & &(2 \pi)^{-3N/2} \int e^{-i \tilde{\bm \pi} \cdot \bm x} f_{KLM}(u,A;\bm x)
 \: d \bm x = \\
 & & (2 \pi)^{-3N/2} \int e^{- \tilde{\bm x} \cdot A \bm x -i \tilde{\bm \pi}
\cdot \bm x} |\tilde{u} \bm x|^{2K} \mathcal{Y}_{LM}(\tilde{u} \bm x) \:
d \bm x. \nonumber
\end{eqnarray}

The principle for the proof is quite similar to what was proposed previously; the
same changes of variables occur; the only difference concerns the integration
over the $\bm Z_1$ variable that must be done now with help of the most general
formula (\ref{eq:TFcorsimp}).

The FT for the general correlated gaussian takes the form
\begin{eqnarray}
\label{eq:FTcasgen}
 & &h_{KLM}(u,A;\bm \pi) = \frac{(-i)^L\, K! \, (\tilde{u}A^{-1}u)^K}{2^{L+3N/2}
(\det A)^{3/2}}  \\
 & & \times L_K^{L+1/2}\left(\frac{(\tilde{u}A^{-1}\bm \pi)^2}{4\tilde{u}A^{-1}u}\right)
 e^{-\frac{1}{4}\tilde{\bm \pi} \cdot A^{-1} \bm \pi} \mathcal{Y}_{LM}
 (\tilde{u}A^{-1}\bm \pi).\nonumber
\end{eqnarray}
In this case, \textbf{the FT of a general correlated gaussian is not proportional
to a general correlated gaussian}.

However, using the series expansion for the Laguerre polynomial
\begin{equation}\label{}
L_n^a(z) = \sum_{m=0}^n (-1)^m \frac{\Gamma(n+a+1)}{(n-m)!
\Gamma(m+a+1)}\frac{z^m}{m!},
\end{equation}
it follows that \textbf{the FT of a general correlated gaussian is
a linear combination of general correlated gaussians}. Explicitly
we have the property
\begin{equation}
\label{eq:TFsumcorg}
\begin{split}
 &h_{KLM}(u,A;\bm \pi) = \frac{(-i)^L }{(2^{3N/2}\det A)^{3/2}}
 \sum_{K'=0}^K \frac{(-1)^{K'} \, K!}{K'! (K-K')!}   \\
 &\times\frac{(2K+2L+1)!!}
{(2K'+2L+1)!!} \frac{(\tilde{u}A^{-1}u)^{K-K'}}{2^{K+K'+L}}
f_{K'LM}(A^{-1}u,\frac{A^{-1}}{4};\bm \pi).
\end{split}
\end{equation}

\section{Relativistic kinetic energy}
\label{sec:Tr}
In many cases, the potential, or any operator, depends on the momenta $\bm
\pi_i = -i \partial/\partial \bm x_i$. If this dependence occurs through an
integer power of the momenta, the method explained above, for example in
section \ref{sec:Tnr}, applies. However this is not always the case.
For such a delicate situation, there is only one remedy : to work in momentum
representation.

Let us suppose that the operator $\hat{O}$ depends on an argument
of type $\tilde{w} \bm \pi$. Then the calculation of the matrix
element $\langle \psi_{K'LM}(u',A') | \hat{O} | \psi_{KLM}(u',A')
\rangle$ = $\langle h_{K'LM}(u',A';\bm \pi) | \hat{O}(\tilde{w}
\bm \pi) | h_{KLM}(u,A;\bm \pi) \rangle$ can be performed using
the expansion (\ref{eq:TFsumcorg}). The dynamical element to be
calculated has a form like
\begin{equation}
\langle f_{K'LM}({A'}^{-1}u',\frac{{A'}^{-1}}{4};\bm \pi) |
\hat{O}(\tilde{w} \bm \pi) | f_{KLM}(A^{-1}u,\frac{A^{-1}}{4};\bm
\pi) \rangle.
\end{equation}
This last element is then computed by application of Eq. (\ref{eq:Vcentgen3})
(or Eq.(\ref{eq:Vcentgen})) with the trivial changes $u \rightarrow A^{-1}u$,
$A \rightarrow A^{-1}/4$, $u' \rightarrow {A'}^{-1}u'$, $A' \rightarrow
{A'}^{-1}/4$.

As an example of application, we consider the relativistic kinetic energy
operator
\begin{equation}
\label{eq:defTr}
T_R = \sum_{i=1}^{N+1} \sqrt{\bm p_i^2 + m_i^2}.
\end{equation}

The separation of the center of mass motion cannot be done properly in this
case. The usual way to consider the intrinsic motion for the system is to
work in the center of mass frame (which is well defined even in the special
case where all particles have a vanishing mass); this means that we put
$\bm P = \bm 0$ in the subsequent formulae. In particular the momentum
$\bm p_i$ relative to the particle $i$ writes (see relation (\ref{eq:pvspi}))
\begin{equation}
\label{eq:momparti}
\bm p_i = \sum_{j=1}^N U_j^{(i)} \bm \pi_j = \tilde{U}^{(i)} \bm \pi
\end{equation}
with
\begin{eqnarray}
\label{eq:coefUij}
U_j^{(i)} & = & 0 \quad \textrm{if } j < i-1, \nonumber \\
U_j^{(i)} & = & -1 \quad \textrm{if } j = i-1, \\
U_j^{(i)} & = & \frac{m_i}{m_{12\ldots j}}\quad \textrm{if } j \geq i.
\nonumber
\end{eqnarray}

In order to present simple formulae, let us focus on the
special case $K=K'=0$. Then
\begin{eqnarray}
\label{eq:TrK0}
 & & \left\langle \psi_{0LM}(u',A') | T_R | \psi_{0LM}(u,A)\right\rangle =
\\
 & &\sum_{i=1}^{N+1} \left\langle h_{0LM}(u',A';\pi) | \, \sqrt{|\tilde{U}^{(i)}
\bm \pi|^2 + m_i^2} \, | h_{0LM}(u,A;\pi) \right\rangle.\nonumber
\end{eqnarray}
The computation of the matrix element in the summation is performed first
using Eq. (\ref{eq:FTcasK=0cg}) and then Eq. (\ref{eq:Vcentpart2}) with the
appropriate parameters.

The final result looks like
\begin{eqnarray}
\label{eq:enercinrel}\nonumber
 & &\left\langle \psi_{0LM}(u',A') | T_R | \psi_{0LM}(u,A)\right\rangle =
 \frac{\mathcal{N}_L}{\sqrt{4 \pi}} \sum_{i=1}^{N+1} Z_i^{-3/2}  \\
  & & \times \sum_{n=0}^L \frac{L!}{2^n (2n+1)!(L-n)!} \\ &&
  \times
 \mathcal{F}\left(2n+2,\frac{1}{4Z_i},m_i\right) \left(\frac{\gamma_i \gamma'_i}{\rho}
 \right)^n \left(1 - \frac{2 Z_i \gamma_i \gamma'_i}{\rho}\right)^{L-n}.
\nonumber
\end{eqnarray}

Again, the matrix element is proportional to the overlap $\mathcal{N}_L$.
The $\rho$ parameter is defined as in formula (\ref{eq:defqqprho}) while
\begin{align}
Z_i&=\tilde{U}^{(i)}A'B^{-1}AU^{(i)}; & \gamma_i &=
\frac{\tilde{U}^{(i)} A'B^{-1}u}{Z_i}; \nonumber\\
\gamma'_i &= \frac{\tilde{U}^{(i)}AB^{-1}u'}{Z_i}
\end{align}
and the integral for the potential is given by
\begin{equation}
\label{eq:defFTr}
\mathcal{F}(k,A,m)=\int_0^\infty du \: e^{-Au^2} u^k \sqrt{u^2+m^2}.
\end{equation}

In fact, we need in Eq. (\ref{eq:enercinrel}), the values of
$\mathcal{F}(k,A,m)$ for even values of $k \geq 2$. We calculated
this function in the following way. Let us define for simplicity
$\mathcal{I}_l = \mathcal{F}(2l+2,A,m)$ and introduce the
parameter $\beta=Am^2/2$. The first values can be found
analytically
\begin{subequations}
\label{eq:premvF}
\begin{eqnarray}
 \mathcal{I}_0  &=& \frac{1}{2A^2} \beta e^{\beta} K_1(\beta) \\
  \mathcal{I}_1 &=&\frac{1}{2A^3} \beta e^{\beta} [\beta
K_0(\beta) + (2 - \beta) K_1(\beta)]
\end{eqnarray}
\end{subequations}
where $K_n(z)$ is the modified Bessel function. The integral for higher $l$
can be obtained by the following recursion formula
\begin{equation}
\label{eq:highvF}
\mathcal{I}_l = \frac{1}{A} \mathcal{I}_{l-1}+\frac{(2l-1)(l+1-2 \beta)}{A^2}
\mathcal{I}_{l-2}.
\end{equation}

All the above formulae allows to get a very efficient way to calculate the matrix
element of the relativistic energy operator for the few-body problem expressed
in terms of correlated gaussians.

\smallskip

\section{Concluding remarks}
\label{sec:rem}
In this paper, we propose a number of new formulae concerning correlated
gaussians that cannot be found in SV. Our formulation deals with general
correlated gaussians (arbitrary number of particles, arbitrary orbital angular
momentum, arbitrary value of the quantum number $K$), but is limited to the
natural parity states and to operators that do not mix spin and space degrees of
freedom.

We present new formulae relative to central potentials. Instead of integrals
containing Hermite polynomials, we proposed new formulae with simpler integrals
which, most of the time, can be evaluated analytically. Moreover, we derived
another formulation for the matrix elements which is more efficient than that
presented in SV. All these formulae can be simplified a lot in the special
case $K = K' = 0$.

But the most interesting point of this paper is the derivation of the Fourier
transform of correlated gaussians. In the peculiar case $K = 0$, we showed that
the Fourier transform is just proportional to a correlated gaussian, but with
renormalized parameters. In the general case, we showed that the Fourier
transform is a combinaison of general correlated gaussians.

These new formulae allow the evaluation of the matrix elements of a
relativistic kinetic energy operator. Since this kind of operator is more and
more introduced in realistic situations, the results of this paper are of
primordial importance for future few-body calculations.

\acknowledgments \label{ackno} We are indebted to C. Semay for a
careful reading of our manuscript, and for fruitful remarks.

\end{document}